# Oxygen Vacancy Induced Flat Phonon Mode at FeSe /SrTiO$_3$ interface

Yun Xie[1], Hai-Yuan Cao[1], Yang Zhou[1], Shiyou Chen[2], Hongjun Xiang[1], and Xin-Gao Gong[1]

[1]State Key Laboratory of Surface Physics, Key Laboratory for Computational Physical Sciences (MOE), Department of Physics, Fudan University, Shanghai 200433, China, [2]Key Laboratory of Polar Materials and Devices (MOE), East China Normal University, Shanghai 200241, China

Email: xggong@fudan.edu.cn

A high-frequency optical phonon mode of SrTiO$_3$ (STO) was found to assist the high-temperature superconductivity observed recently at the interface between monolayer FeSe and STO substrate. However, the origin of this mode is not clear. Through first-principles calculations, we find that there is a novel polar phonon mode on the surface layers of the STO substrate, which does not exist in the STO crystals. The oxygen vacancies near the FeSe/STO interface drives the dispersion of this phonon mode to be flat and lowers its energy, whereas the charge transfer between STO substrate and FeSe monolayer further reduces its energy to 81 meV. This energy is in good agreement with the experimental value fitted by Lee *et al.* for the phonon mode responsible for the observed replica band separations and the increased superconducting gap. The oxygen-vacancy-induced flat and polar phonon mode provides clues for understanding the origin of high Tc superconductivity at the FeSe/STO interface.

Recently the extremely enhanced superconductivity in monolayer FeSe grown on Nb-doped SrTiO$_3$ (STO) substrate have been reported[1-15]. The signature of Tc higher than the boiling point of the liquid nitrogen (77 K)[1-2,4-8] has also been revealed in this system, which is considerable higher than that of any other iron-based superconductor[16,17]. Since the bulk FeSe exhibits a Tc only about 8 K[18,19], the extremely high Tc of single-layer FeSe/STO system indicates a critical role of the interactions between the single FeSe layer and STO substrate[1]. The very recent work revealed a direct correlation between the interfacial phonon and high Tc superconductivity. In monolayer FeSe on STO substrate, each energy band of the FeSe film is exactly replicated at a fixed energy separation, which is observed by ARPES[2]. The energy separation between the replica and main band was very likely to be a shake-off satellite band induced by an interfacial optical phonon mode of STO substrate. The interaction between FeSe electrons and the particular STO phonon mode is also believed to strengthen the Cooper pairing[20] and be responsible for enlarging the superconducting gap in the single-layer FeSe/STO system.

It is so far not clear how the FeSe monolayer and the STO phonon mode coupled with each other, or why the experimentally observed replica bands appear. One possibility could be from the well-separated phonon mode of bulk STO[21], whose energy is around 100 meV, closed to the experimental replica band separation. However, if it was true, what was the role of surface or interface phonon. Otherwise, we need to identify the origin of the specific phonon mode near the FeSe/STO interface, which is crucial not only for understanding the mechanism of the interfacial superconductivity but also for further increasing Tc. Although the phonons of bulk STO have been widely studied, those of the surface of STO substrate have seldom been systematically studied. In this paper, based on the first principles calculations, we have carefully studied the vibrational properties of the FeSe/STO interface, and successfully identify a special phonon mode, which is strongly related to the oxygen vacancy on STO surface and the charge transfer between the FeSe monolayer and STO substrate. The calculated energy of the phonon mode is in accordance with the recent observed particular surface phonon mode, which is very different from bulk STO.

**Methods**

The calculations were performed in the framework of density functional theory (DFT) with the local density approximation (LDA) as implemented in the Vienna ab initio simulation package (VASP)[22,23] code with projector augmented wave (PAW) pseudopotentials[24,25]. We use a Hubbard U term[26] with U=3.2 eV and J=0.9 eV for the Ti 3d states in all our calculations. The calculations of phonon dispersion are based on density functional perturbation theory (DFPT)[27-29]. An energy cutoff of 550 eV was adopted in all the calculations. The Brillouin zone was sampled with a K point mesh of 6×6×1 in the structure relaxation and the DFPT calculations. All the atomic coordinates were relaxed until the atomic forces are smaller than 0.01 eV/ Å.

The antiferrodistortive (AFD)[30] phase (I4/mcm) of the bulk STO is considered in the current study, and the original unit cell contains 20 atoms as each pair of neighboring $TiO_6$ octahedral rotates in opposite directions. The $TiO_6$ octahedral rotation angle in the initial AFD phase of STO is set to be 2.1 $^{\circ}$[31]. The lattice constant a and b are set to be 5.507 Å, c is set to be 7.797 Å, in accordance with the experimental values[32]. To study the surface and FeSe/STO interface, we constructed $TiO_2$-terminated STO slab model which contains five $TiO_2$ layers and four SrO layers with a vacuum layer as thick as 15 Å. The un-relaxed and relaxed atomic structures of $TiO_2$-terminated tetragonal phase $SrTiO_3$ (001) slabs are shown in Fig. 1. The black arrows in the un-relaxed STO slab represent the AFD rotation below 105 K. The Ti-O bond length between the top two layers of the STO slab becomes shorter after relaxation because of the breaking one Ti-O bond around the surface Ti atom.

The oxygen vacancies play an important role as electrons donors in the superconducting FeSe/STO interface. Several experiments have observed two-dimensional electron gas (2DEG) in both vacuum-cleaved[33,34] and annealed[35] STO surfaces, which were believed to be introduced by the concentrated oxygen vacancies in the surface of STO. The structure with oxygen vacancies on the top layer of STO substrate is also in accordance with the Se etching used to prepare the high-Tc samples of monolayer FeSe/STO[1,38]

To study the effects of the surface defect, we introduce the oxygen vacancies by removing one oxygen atom from the top $TiO_2$ layer in the STO slab. Figure 3(a) gives an example of the relaxed atomic structure of the STO slab with oxygen vacancies (STOov). Such configuration contains oxygen vacancies along in-plane Ti-O-Ti dimmers, each oxygen vacancy locates between two Ti atoms. Each Ti atom in the top $TiO_2$ layer has exactly one neighboring oxygen vacancy. The existence of the oxygen vacancy reduces the interlayer Ti-O bond length between the top two layers of the relaxed STOov slab by as large as 0.34 Å. Such large change of interlayer Ti-O bond length would play an essential role in the appearance of a surface polar phonon mode.

**Results**

First we calculated the phonon spectrum of the cubic STO crystal, as shown in Fig. 2(a). There are two imaginary frequencies near the M point and R point of the Brillouin zone (BZ) that is similar to the previous first-principles calculations[21]. The imaginary phonon mode at R point indicates the AFD phase transition near 105 K. A large phonon band gap appears in the phonon spectrum with the LO-TO splitting, and there is no phonon mode in the energy range 70-90 meV. Above the gap, there is a phonon mode about 97 meV at Γ point. This high-energy mode of bulk STO is obviously separated from all other phonon bands.

**For the STO surface, a surface polar phonon mode appears, which is very different from the isolated phonon mode of bulk STO.** Figure 2(b) shows the phonon spectrum of relaxed STO slab where a new surface polar mode appears with the energy around 107.5 meV at Γ point. This mode is mainly composed of relative Ti and O atomic displacements along (001) direction in the top two layers of the relaxed STO slab. The atoms in the same layer have the same vibrating phase, which means that the induced polar field will not be canceled. This kind of

surface polar phonon mode has not been found neither in the phonon spectrum of bulk STO nor in that of the un-relaxed STO slab. If the STO slab is not relaxed, the interlayer Ti-O bond lengths between the top two layers are same as those in the bulk, then our calculation does not find any new mode around 107.5 meV. After the relaxation of the STO surface structure, the Ti-O bond length becomes shorter, and meanwhile the new phonon mode appears with larger energy.

**As shown in Fig. 4, the oxygen vacancies on the STO surface (STOov) can significantly influence the surface phonon mode.** The oxygen vacancies would shift the energy of the surface polar mode from 107.5 meV to 93.6 meV at Γ point. Furthermore the mode surprisingly becomes almost flat (Fig. 4(b)). A schematic view of the surface polar mode of STOov slab is shown in the left panel of Fig. 3, the purple arrows represent the related atomic vibrating eigenvectors, which indicates that this mode is closely related to localized vibration near the oxygen vacancy, this could be the reason why it is almost flat in the Brillouin zone.

As electron donors, the oxygen vacancy will produce an amount of free electrons on the surface and largely increase the in-plane dielectric constant near the surface layers of STO substrate. The dynamical matrices of the LO and TO modes at q=0 in bulk STO are related by[36]:

$$D_{mn}^{LO} = D_{mn}^{TO} + \frac{4\pi e^2}{\Omega} \frac{Z_m^* Z_n^*}{\varepsilon_\infty(0)} \quad (1)$$

where $\varepsilon_\infty(0)$ represents the static in-plane dielectric constant. Since the in-plane dielectric constant increases, the magnitude of the LO-TO splitting will be reduced, thus the high-energy phonon mode of bulk STO will not exist anymore.

It is well known that the FeSe monolayer on STO with oxygen vacancies was heavily n-type doped which were experimentally observed[11] and theoretically predicted[37,38]. The Hall experiment of the FeSe thin film[4] also demonstrated the n-type property of monolayer FeSe above Tc. The in situ scanning tunneling spectroscopy (STS) results about insulating behavior of the exposed STO surface but superconducting behavior of monolayer FeSe further provide the evidence of charge transfer from STO surface to FeSe layer[3]. Such charge transfer between FeSe and STO can change the atomic structure of the STOov substrate surface, especially the interlayer Ti-O bond length near surface layers and thus the interfacial mode. As shown in Fig. 3(b), the interlayer Ti-O bond length between top two layers of STOov substrate in FeSe/STOov slab is 0.58 Å longer than that in STOov slab, as shown in Fig. 3(a).

**With the monolayer FeSe added on the STO substrate, the interaction between FeSe and STO substrate further decreases the phonon energy of the surface polar mode.** The energy of the surface polar mode decreases to around ~81.7 meV at Γ point (Fig. 4(c)), which is about 12 meV lower than that of the surface polar mode without the FeSe monolayer. It is clear that such surface atomic vibrations can induce dipole in the direction perpendicular to the FeSe/STOov interface, and these polar displacements can create net dipole moments along (001) and exert an electric field on the FeSe monolayer. The electrons in FeSe monolayer are confined in the plane parallel to the interface, so the parallel electric field will be screened but the perpendicular electric field will not. Therefore this polar mode in the direction perpendicular to FeSe/STO interface may have the largest electron-phonon coupling strength. The strength of such electron-phonon coupling is suggested to be enhanced by a factor of $\sqrt{\varepsilon_\parallel/\varepsilon_\perp}$, where $\varepsilon_\parallel$ and $\varepsilon_\perp$ represent the anisotropic dielectric constant parallel and perpendicular to the interface, respectively[2].

The interlayer Ti-O bond length in the top two layers of STO substrate will affect the phonon energy of this surface phonon mode significantly. The changes of the phonon energy of surface polar mode can be qualitatively understood from the Ti-O bond length and the number of neighboring bonds. From the relaxed STO surface without O vacancy to that with O vancancy (STOov), the phonon energy of surface polar mode decreases. The reason is that in the relaxed STO slab there are four interlayer O-O bonds near the Ti-O bond while there are only three in the

STOov slab (Fig. 1(b), Fig. 3(a)). These O-O bonds will strengthen the force constant of the vibrations of the interlayer Ti and O atoms. Although a larger interlayer Ti-O bond length appears in the relaxed STO slab, the increased number of neighboring O-O bonds will further increase the elastic constant and result higher surface phonon energy than in the STOov slab. When the FeSe monolayer is placed on STOov substrate (Fig. 3(a), 3(b)), the electron transfers from STOov slab surface to FeSe monolayer, and the Ti-O bond length increases. A longer Ti-O bond length leads to a smaller Ti-O force constant between them, thus the surface polar mode near the FeSe/STOov interface has lower phonon energy than that of the STOov surface.

Figure 4 shows the phonon dispersion with energies higher than 60 meV for STO, $STO_{OV}$ and FeSe/STOov slabs. Among them the surface polar phonon mode which we have described above are highlighted by red lines. The surface polar mode are located at 107.5 meV, 93.6 meV, 81.7 meV around Γ point in the three slabs, respectively, and it shows a very flat dispersion in the STOov and FeSe/STOov slabs near Γ point.

We have checked all the phonon mode with energy higher than 60 meV, there are two more similar modes, but with much smaller surface dipole moment. We have calculated the ratio of the "surface vibration" in STO, STOov, and FeSe/STOov slabs. The ratio of the surface vibration is calculated by summing up the square of the phonon eigenvectors of the atomic displacements in the top two layers of STO substrate. The ratio of the phonon modes with the phonon energy above 60 meV is shown in Fig. 5. The blue ones represent the ratio of atomic vibrations in the top two layers of STO substrate and the red ones represent the ratio of the atomic vibration which can induce surface dipole along (001) direction. Though there are several phonon modes composed mainly of the surface atomic vibrations (blue ones), only one mode is dominated by the surface polar vibrations (red ones) in the STOov and FeSe/STOov slabs.

**Discussion**

Recently, Lee *et al.*[2] observed the band replica in the monolayer FeSe/STO which was attributed to a high-frequency optical phonon mode of the STO surface. They proposed that the interaction between the FeSe electrons and a particular phonon mode from STO substrate is also responsible for raising the superconducting gap opening temperature in the monolayer FeSe/STO system. They used a dispersionless phonon mode of 80 meV to fit their experimental data and got a very good agreement. However, the origin of this special mode is not clear. Based on our calculated results, we predict that the oxygen-vacancy induced flat phonon mode with the energy of 81.7 meV in FeSe/STO slab is very likely to be the interfacial mode proposed experimentally. According to our calculations, this flat surface phonon mode is unique, which is also consistent with the experimental phenomenon that only one replica band can be effectively observed. Furthermore our results are also in good agreement with the results of Peng *et al.*[7], who found that the energy separation between main band and the replica band from various oxide substrates are all around 90 meV.

In conclusion, we have systematically investigated the phonon modes of the STO surface, and found a special surface polar phonon mode which is mainly composed of the Ti and O atomic vibrations perpendicular to the surface. This mode becomes flat and dispersionless when there are oxygen vacancies on the surface. With the FeSe monolayer deposited on the surface, the strong interaction between FeSe monolayer and STO surface significantly decreases the energy of the phonon mode to lower value at 81.7 meV, which is consistent with the interfacial phonon mode observed recently near the FeSe/STO interface with high-temperature superconductivity[2]. As a result of the surface dipole field induced by the surface mode, the electron-phonon coupling can be strong near the interface, which might play an important role in the superconductivity observed by Lee *et al.*[2] near the FeSe/STO interface.

**Acknowledgements**

We acknowledge professor Z.-X. Shen for stimulating discussions. The work was partially supported by the Special Funds for Major State Basic Research, National Natural Science Foundation of China (NSFC). Computation was performed at the Supercomputer Center of Fudan University.

**Author contributions**

Y. X. and H.-Y.C. contributed to first-principles calculations and article writing.

Y.Z., S.-Y.C., H.-J.X. contributed to the discussion of results and manuscript refinement. X.-G.G. suggested the calculations and contributed significantly to the discussion of results and manuscript refinement.

**Additional Information**

**Competing financial interest**: The authors declare no competing financial interest.

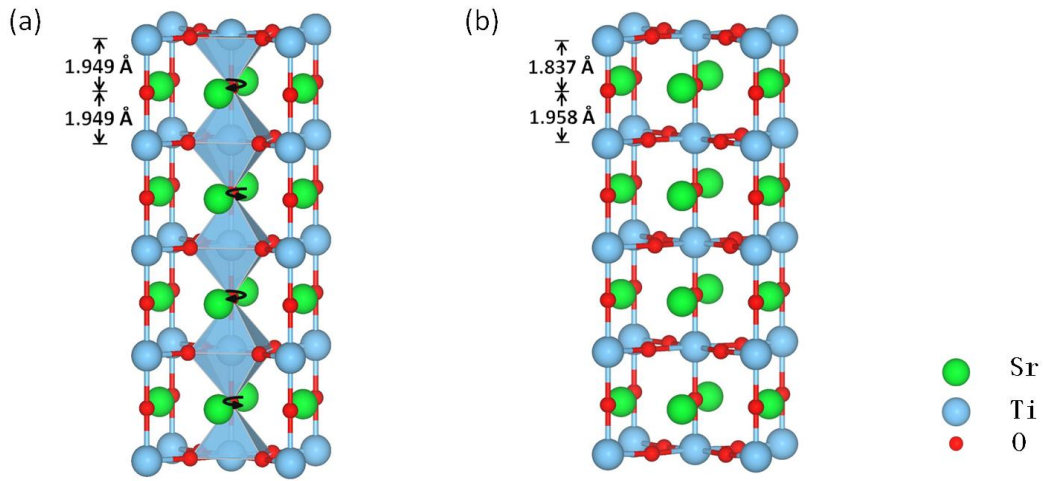

**Figure 1 | The atomic structures of (a) the unrelaxed and (b) relaxed TiO$_2$ terminated tetragonal SrTiO$_3$ slab.** The black arrows indicated the antiferrodistortive rotation in tetragonal phase SrTiO$_3$ below 105 K, and the interlayer Ti-O bond length in top three layers are denoted.

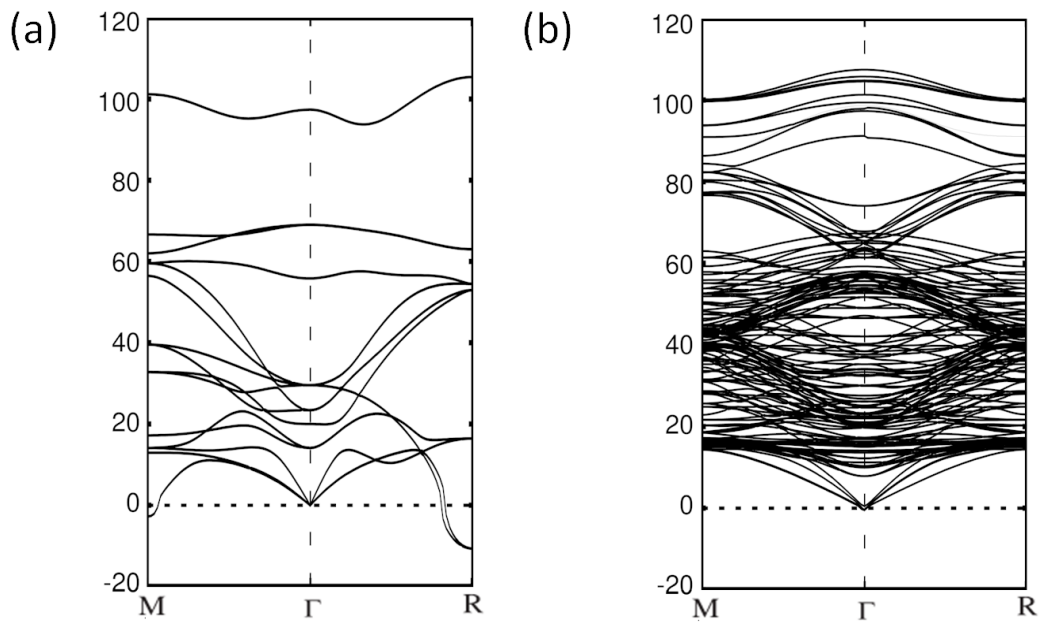

**Figure 2 | The calculated phonon dispersion along high symmetry q-points.** (a) Cubic SrTiO$_3$. (b) Relaxed tetragonal SrTiO$_3$ (001) slab. An isolated phonon mode appears in the phonon spectrum of cubic SrTiO$_3$ with a large phonon gap (70 meV – 90 meV).

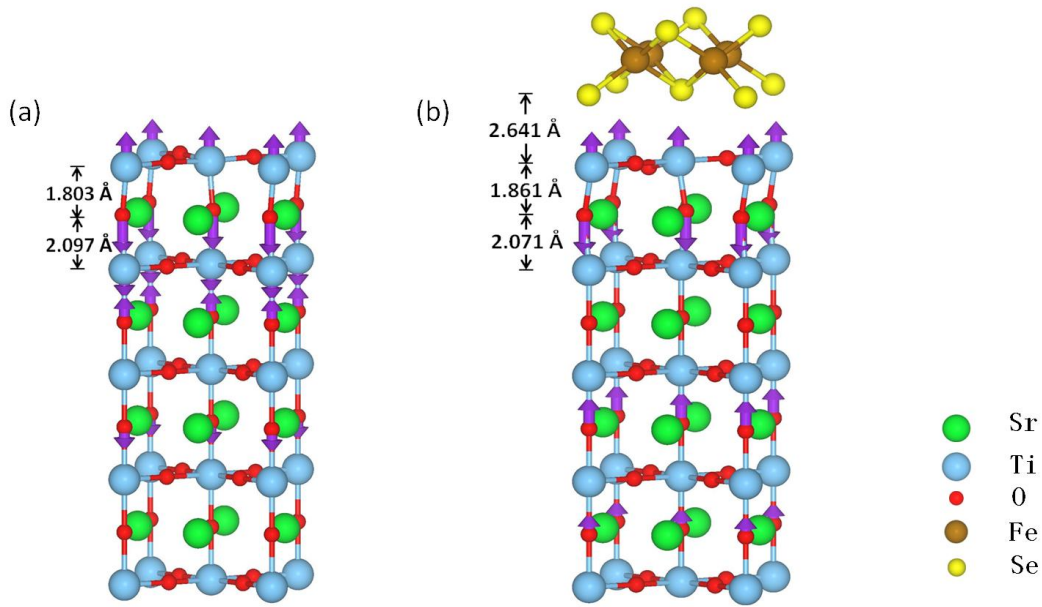

**Figure 3 | Illustration of the atomic displacements of the surface polar phonon mode.** (a) Relaxed $STO_{ov}$ (001) slab. (b) Relaxed atomic structure of FeSe monolayer on $STO_{ov}$ (001) slab. This particular phonon mode is mainly composed of the polar vibrations near STO surface.

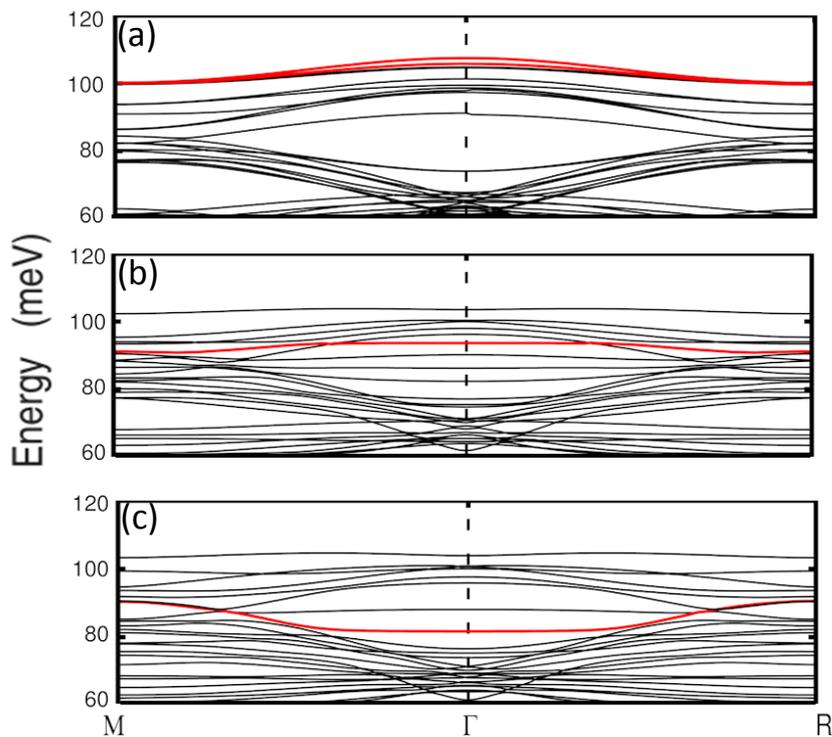

**Figure 4 | Calculated phonon dispersion only showing the mode with energy larger than 60 meV.** (a) Relaxed STO (001) slab. (b) Relaxed $STO_{ov}$ (001) slab. (c)

relaxed FeSe monolayer on STO$_{ov}$ (001) slab. The phonon modes which can induce the surface dipole along (001) direction are highlighted by red lines.

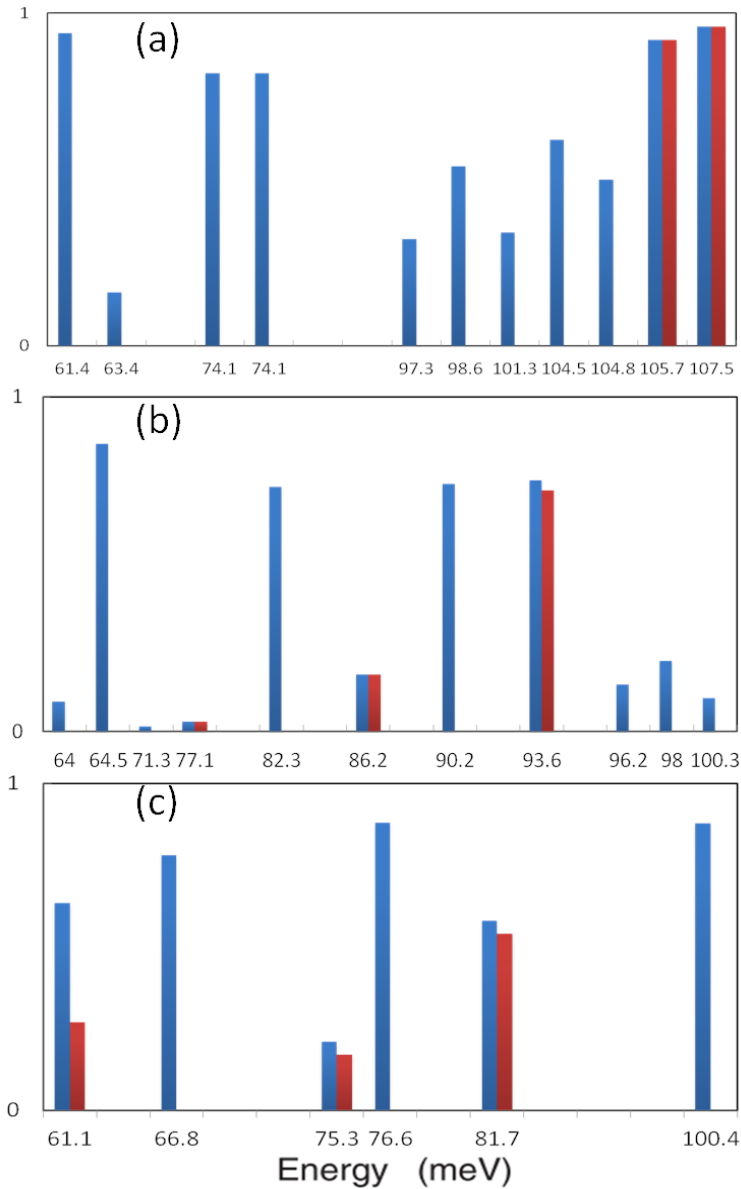

■The ratio of atomic vibration in the top two layers of STO$_{ov}$ substrate

■The ratio of atomic vibration in the top two layers of STO$_{ov}$ substrate which can induce dipole along (001) direction

**Figure 5 | The ratio of surface vibration (blue) and surface polar vibration (red).** (a) STO(001) slab, (b) STO$_{ov}$ (001) slab, and (c) FeSe monolayer on STO$_{ov}$ (001) slab. The phonon modes of (b) 93.6 meV and (c) 81.7 meV are dominated

by the surface polar vibration.